\documentclass[aps,showpacs,english,preprint,nofootinbib]{revtex4}
\usepackage{mathrsfs}
\usepackage{graphicx}
\usepackage{amsmath, amssymb}
\usepackage{babel}
\usepackage{color}

\def\lsim{\mathrel{\rlap{\lower4pt\hbox{\hskip1pt$\sim$}}
    \raise1pt\hbox{$<$}}}         
\def\gsim{\mathrel{\rlap{\lower4pt\hbox{\hskip1pt$\sim$}}
    \raise1pt\hbox{$>$}}}         

\begin{document}

\hfill NPAC-12-15

\title{Parity Violating Deep Inelastic Electron-Deuteron Scattering: Higher Twist and Parton Angular Momentum}

\author{Chien-Yeah Seng$^{a}$}
\author{Michael J. Ramsey-Musolf$^{a,b}$}

\affiliation{$^{a}$University of Wisconsin-Madison, Madison,
Wisconsin 53706, USA} \affiliation{$^{b}$California Institute of
Technology, Pasadena, California 91125, USA}

\date{20 February 2013}

\begin{abstract}

We study the effect of parton angular momentum
 on the twist-four correction to the left-right asymmetry in
the electron-deuteron parity-violating deep inelastic scattering
(PVDIS). We show that this higher-twist correction is transparent to
the dynamics of parton angular momentum needed to account for the
Sivers and Boer-Mulders functions and spin-independent parton
distribution functions. A sufficiently precise measurement of the
PVDIS asymmetry may, thus, provide additional information about the
parton dynamics responsible for nucleon spin.

\end{abstract}


\maketitle

\section{Introduction}

As a complement to the studies at high-energy frontier, measurements at the intensity frontier
(or precision frontier) provide powerful tools in the
search for physics Beyond the Standard Model (BSM). Observables such
as the muon anomalous magnetic moment are measured to very high
precision, and experimental results are then compared with
theoretical predictions. To the extent that the latter are
sufficiently reliable, any possible deviation would point to BSM
physics. Alternately, these experiments can provide
new insights into the dynamics of the Standard Model.

Electron-deuteron parity violating deep inelastic scattering ($e$D
PVDIS) is an excellent example of this class of studies.
Historically, it provided the first experimental measurement of weak
mixing angle $\theta_W$ \cite{Prescott}. Nowadays, with the prospect
of the Jefferson Laboratory 12-GeV upgrade and the use of a new
spectrometer called SoLID, the left-right asymmetry of PVDIS can be
measured with 0.5\% precision over the kinematic range $0.3<x_B<0.7$
\cite{JLab}. With this level of precision, one will be able to probe
or constrain an interesting set of BSM scenarios, such as a
leptophobic  Z' boson\cite{Buckley:2012tc,Martin} and
supersymmetry\cite{Kurylov:2003xa}, as well as to study  hadronic
physics effects which are yet to be fully understood, such as charge
symmetry violation (CVC) and higher-twist (HT).

The effect of HT \cite{Jaffe_lecture} is a potentially important
Standard Model correction that originates from the interaction
between partons. This correction in general scales as
$(Q^2)^{-(\tau-2)/2}$, with the twist $\tau>2$, so its effect is
enhanced at low $Q^2$. In the framework of the operator product
expansion (OPE), the higher-twist correction can be expressed as a
convolution of a high-energy and low-energy piece; the former
(embodied in the Wilson coefficients) can be calculated using
perturbative methods, whereas the latter involves hadronic matrix
elements that require understanding of non-perturbative QCD.
Studying the higher-twist correction may help us in probing
correlations between the confined quarks and gluons inside the
nucleon, so it is interesting to explore HT matrix elements within
various model approaches. One advantage of $e$D PVDIS process is
that the HT contribution to the leading term in the PV asymmetry
(defined below) arises from a single operator matrix element and
can, in principle, be separated kinematically from the subleading
terms that have a more complicated HT structure. With this
motivation in mind, several previous works
\cite{Castorina:1985uw,Sonny,Belitsky,Hobbs} have been carried out
to study the twist-four (i.e. $\tau=4$) correction to the left-right
asymmetry of $e$D PVDIS. In what follows, we report on a study that
follows-up these earlier works.

The study of HT may also shed light on another important issue,
namely, the spin structure of the nucleon. Nearly twenty-five years
ago, the EMC collaboration \cite{EMC} performed a DIS experiment
with longitudinally-polarized muons on a target of
longitudinally-polarized protons, obtaining a value for the
structure function $g_1(x_B)$ over the range $0.01 < x_B < 0.7$.
After extrapolating to the low- and high-$x_B$ region, the
collaboration obtained a value for the leading moment of  $g_1(x_B)$
that contradicted the Ellis-Jaffe sum rule \cite{Ellis_Jaffe} and
implied that that the spin of proton is not built up entirely from
the quark spin. The result has been confirmed by a variety of
subsequent studies. A key question in nuclear physics research has,
thus, become explaining in detail the source of nucleon spin in
terms of QCD degrees of freedom.

From a theoretical perspective, arriving at a decomposition of the
nucleon spin in terms of gauge-invariant matrix elements of local
operators that afford a straightforward partonic interpretation has
been a vexing problem, and different approaches have been pursued
over the years\cite{Burkardt,JaffeOAM,Ji1207,Ji78,Ji109}. In each
case, reference is usually made to the interpretation in the
light-cone -- gauge dependence notwithstanding --  given its
historical importance for thinking about parton dynamics. However,
while the meaning of the quark helicity is gauge invariant, the
relative importance of other aspects of partonic angular momentum
(gluon helicity and quark and gluon orbital angular momentum) in
general vary with the choice of gauge and even definition.
Nonetheless, it is interesting to ask how different observables may
probe different aspects of partonic angular momentum and to do so in
a way that is both gauge-invariant and as insensitive as possible to
a particular angular momentum decomposition.

In this respect, we will study HT in the context of light-cone quantization. In early work within this framework, it has been shown that one particular component of parton angular momentum -- identified as quark orbital angular momentum (OAM) under
light-cone quantization using light-cone gauge -- is responsible
for the non-zero value of Sivers function and Boer-Mulders function
\cite{Pasquini_1,Pasquini_2} in semi-inclusive deep inelastic
scattering (SIDIS) \cite{Prokudin}. In light of these results, it is
also interesting to study how the inclusion of the same
component of parton angular momentum modifies the current model
predictions for HT corrections to $e$D PVDIS. Indeed, in all the previous studies of $e$D PVDIS, only the Fock component of the nucleon wavefunction with
zero parton OAM has been included.

After including quark OAM in the light-cone amplitudes,
we observe a rather non-intuitive phenomenon:
although the absolute magnitude of individual
non-zero quark OAM contributions can be significant,
they largely cancel against each other. We will argue that this
cancelation is largely independent of the detailed model for the relevant light-cone amplitudes. As a result, the
twist-four correction to PVDIS is almost transparent to the
inclusion of quark OAM. In contrast, other hadronic quantities,
such as the parton distribution functions (PDF), Sivers function,
and Boer-Mulders function, manifest non-negligible dependence on
quark OAM. Generalizing from the particular choice of light-cone quantization and light-cone gauge, we thus conclude
that whatever features of parton angular momentum are responsible for the observed behavior of the PDFs, Sivers, and Boer-Mulders functions,
they should have a relatively minor impact on the HT correction to $e$D PVDIS of interest here. Moreover, any deviation from the light-cone predictions obtained  here and
in previous works\cite{Castorina:1985uw,Sonny,Belitsky} -- should they be observed expermentally -- would signal the importance of other aspects of parton angular momentum and/or higher Fock space components of the nucleon wavefunction.

The discussion of the computation leading to these observations  is
arranged in the following order: in Section II we summarize the
relevant results of the general formulation of the twist-four
correction to $e$D PVDIS; in Section III we introduce the light-cone
wavefunction with quark OAM-dependence; in Section IV we present the
analytic expressions of the hadronic matrix elements needed for the
twist-4 correction, and demonstrate the generic cancelation between
non-zero quark OAM components; in Section V we present the numerical
results using one specific choice of nucleon wavefunction, and
discuss their physical significance. Detailed formulae appear in the
Appendix.

\section{Higher-twist in PVDIS:  general formulation}

\begin{figure}
\includegraphics[scale=0.3]{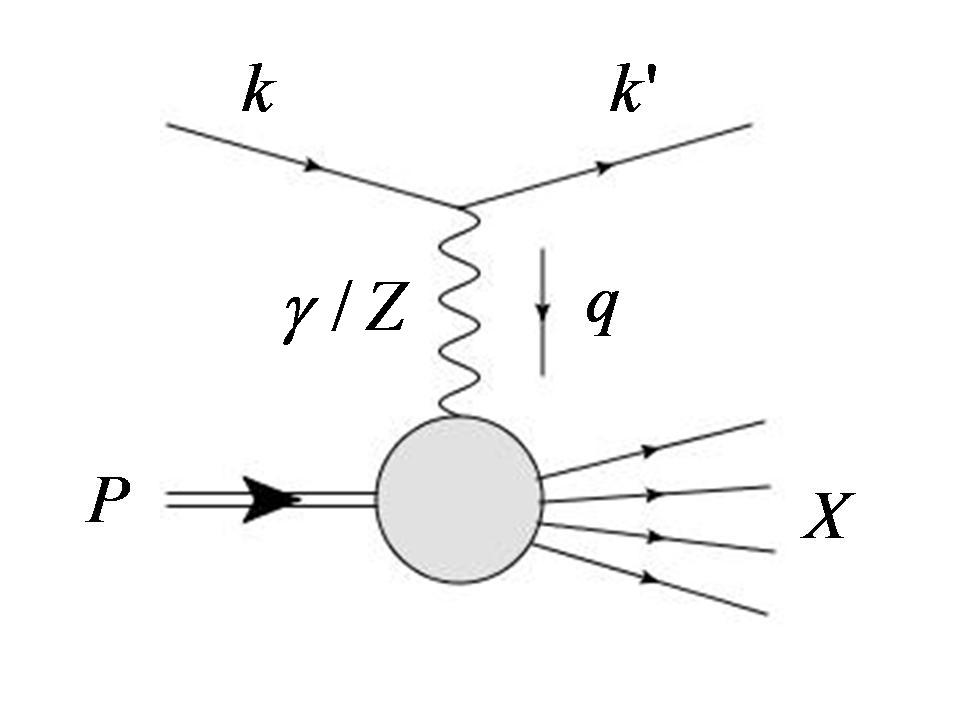}\caption{\label{fig:DIS}Kinematics of e-D PVDIS: a deuteron of momentum $P$ interacts with an incoming electron of momentum $k$ via an exchange
of a single photon or Z-boson, and breaks into hadrons which are
denoted collectively as $X$}
\end{figure}

Here, we review the well-known results for the twist-four correction
in $e$D PVDIS. We will simply quote the central equations that are
relevant to our study without any derivation and refer the reader to
Refs.~\cite{Sonny,Belitsky} for the details.

In $e$D PVDIS, longitudinally-polarized electron beams are incident
on unpolarized deuteron targets. One measures the PV right-left
asymmetry
\begin{equation}
A_{RL}=\frac{d\sigma_R-d\sigma_L}{d\sigma_R+d\sigma_L}
\end{equation}
where $d\sigma_{R/L}$ is the differential cross-section for the
scattering of the right/left-handed electrons. At the one-boson exchange (OBE) level,
the leading parity-violating piece comes from the interference
between photon and Z-boson exchange diagrams (see Fig \ref{fig:DIS}). The low-energy Z-boson
exchange interaction can be described by the following effective
4-fermion interaction:
\begin{equation}
\mathcal{L}_{PV}=\frac{G_F}{\sqrt{2}}[\bar{e}\gamma^\mu\gamma_5
e(C_{1u}\bar{u}\gamma_\mu u+C_{1d}\bar{d}\gamma_\mu
d)+\bar{e}\gamma^\mu e(C_{2u}\bar{u}\gamma_\mu\gamma_5
u+C_{2d}\bar{d}\gamma_\mu\gamma_5 d)]\end{equation}
where, at tree level, we have:
\begin{eqnarray}
C_{1u}&=&-\frac{1}{2}+\frac{4}{3}\mathrm{sin}^2\theta_W\label{eq:tree1}\\
C_{1d}&=&\frac{1}{2}-\frac{2}{3}\mathrm{sin}^2\theta_W\label{eq:tree2}\\
C_{2u}&=&-\frac{1}{2}+2\mathrm{sin}^2\theta_W\label{eq:tree3}\\
C_{2d}&=&\frac{1}{2}-2\mathrm{sin}^2\theta_W\label{eq:tree4}
\end{eqnarray}
Neglecting contributions from sea quarks, assuming charge symmetry
($u_V^p=d_V^n$, {\em etc.} with $q_V^N$ being the valence quark PDF
of nucleon $N$), the leading-twist SM prediction
 is given by the Cahn-Gilman formula\cite{Cahn:1977uu}:
\begin{equation}A_{RL}=\frac{G_F
Q^2}{2\sqrt{2}\pi\alpha}\frac{3}{5}[(2C_{1u}-C_{1d})+(2C_{2u}-C_{2d})\frac{1-(1-y)^2}{1+(1-y)^2}]\end{equation}
where $Q^2=-q^2$ and $y=P\cdot q/P\cdot k$.

To include corrections from possible BSM and as well as other SM
pieces, we can reparametrize the Cahn-Gilman formula \cite{Sonny}:
\begin{equation}A_{RL}=-\frac{G_F
Q^2}{2\sqrt{2}\pi\alpha}\frac{3}{5}[\tilde{a}_1+\tilde{a}_2\frac{1-(1-y)^2}{1+(1-y)^2}]\end{equation}
with $\tilde{a_i}=-(2C_{iu}-C_{id})(1+R_i)$. Here, $R_i$ describes any deviation of the $C_i$ from the expressions in Eqs.~\eqref{eq:tree1} to \eqref{eq:tree4},
including both SM and BSM corrections. In this paper we concentrate on $R_1^{HT}$, namely the higher-twist correction to $\tilde{a}_1$.

Bjorken and Wolfenstein \cite{Bjorken,Wolfenstein} showed that, if
one assumes isospin symmetry and neglects sea quark contributions, then there
is only one matrix element that contributes to $R_1^{HT}$ (for a detailed review of these arguments in a more
modern context, see Ref.~\cite{Sonny}). This
observation significantly simplifies the theoretical interpretation of the asymmetry, allowing us to
concentrate on one particular matrix element without needing to
to disentangle the contributions from many different sources. In brief, the
Bjorken and Wolfenstein argument works as follows: $A_{RL}$
arises from the interference between the electromagnetic and weak neutral
currents. First, one can decompose both currents into an isoscalar $S$
and an isovector $V$ term. The matrix elements of the
$S\times V$ cross-term vanishes because deuteron is an isosinglet.
Furthermore, at leading twist, we have $\left\langle
SS\right\rangle= \left\langle VV\right\rangle$. Therefore, the
difference between $ \left\langle SS\right\rangle$ and $\left\langle
VV\right\rangle$ that enters hadronic tensor $W_{\mu\nu}$
\begin{equation}
W_{ud}^{\mu\nu}(P,q)=\frac{1}{8\pi M_D}\int d^4ze^{iq\cdot
z}\left\langle D(P)\right|\bar{u}(z)\gamma^\mu
u(z)\bar{d}(0)\gamma^\nu d(0)+(u\leftrightarrow
d)\left|D(P)\right\rangle\label{eq:HT_matrix}
\end{equation}
with $M_D$ being the mass of deuteron, is the only matrix element
giving a HT correction to $R_1$.

Below, we will compute the matrix element \eqref{eq:HT_matrix} using an
expansion of string operators \cite{Balitsky_Braun}  in
order to extract the twist-four piece; the latter is expressed in terms
of the deuteron twist-four distribution function $ \tilde{Q}_D(x_B)$, which will be computed in Section IV.

\section{The light-cone amplitudes}

The main challenge in proceeding from \eqref{eq:HT_matrix} is our
ignorance of the details of the nucleon wavefunctions. As QCD is
non-perturbative at the hadronic scale,  analytical expressions for
the wavefunctions are unknown. At present, lattice QCD can provide
only HT contributions to structure function moments and not the
$x_B$-dependence of the $R_1^{HT}$ that is of interest to the SoLID
experiment. Consequently,  one must turn to  various models that
seek to incorporate non-perturbative dynamics. Previous works on
$R_1^{HT}$  include the use of MIT bag model \cite{Sonny} and
isotropic light-cone wavefunctions that contain both quark and gluon
Fock components \cite{Belitsky}; their results yield similar shapes
for the $x_B$-dependence but differ somewhat in magnitude, with a
maximum $R_1^{HT}$ of $0.003\sim0.005$ at $0.2<x_B<0.7$ for
$Q^2=4\mathrm{GeV}^2$, which is a little bit lower than the
achievable precision level in the SoLID experiment.

In this work we study how the inclusion of additional parton angular
momentum might modify the $R_1^{HT}$ prediction. For this purpose,
we adopt the formalism developed in Ref.~\cite{Ji}, starting from a
light-cone formulation of quark states which is equivalent to the
well-known ``infinite momentum frame" point of view that gives the
PDF its intuitive meaning as a parton momentum probability
distribution \cite{Jaffe_lecture}. We then perform  a light-cone
expansion of the nucleon state, retaining only the portion of Fock
space containing three valence quarks with all possible quark OAM.
To illustrate, we consider a spin-up proton. Its three valence
quarks can form a total helicity of $\pm1/2,\pm3/2$; therefore in
order to keep the total proton spin in z-direction to be 1/2 we need
to assign different z-component quark OAM (i.e. $l_z$) for different
combinations.

A spin-up proton state, then, can be parametrized as the follows:
\begin{equation}\left|P\uparrow\right\rangle=\left|P\uparrow\right\rangle^{l_z=0}+\left|P\uparrow\right\rangle^{l_z=1}+\left|P\uparrow\right\rangle^{l_z=-1}+\left|P\uparrow\right\rangle^{l_z=2}
\end{equation}
with
\begin{eqnarray}
\left|P\uparrow\right\rangle^{l_z=0}&=&\frac{\epsilon^{abc}}{\sqrt{6}}\int[DX_3](\psi^{(1)}(1,2,3)+i(k_1^xk_2^y-k_1^yk_2^x)\psi^{(2)}(1,2,3))\times
\nonumber\\
&&u_{a\uparrow}^\dagger(1)\{u_{b\downarrow}^\dagger(2)d_{c\uparrow}^\dagger(3)-d_{b\downarrow}^\dagger(2)u_{c\uparrow}^\dagger(3)\}\left|0\right\rangle\label{eq:LC1}\end{eqnarray}
\begin{eqnarray}\left|P\uparrow\right\rangle^{l_z=1}&=&\frac{\epsilon^{abc}}{\sqrt{6}}\int[DX_3](k_{1\perp}^+\psi^{(3)}(1,2,3)+k_{2\perp}^+\psi^{(4)}(1,2,3))\times\nonumber\\
&&(u_{a\uparrow}^\dagger(1)u_{b\downarrow}^\dagger(2)d_{c\downarrow}^\dagger(3)-d_{a\uparrow}^\dagger(1)u_{b\downarrow}^\dagger(2)u_{c\downarrow}^\dagger(3))\left|0\right\rangle\label{eq:LC2}\end{eqnarray}
\begin{eqnarray}\left|P,\uparrow\right\rangle^{l_z=-1}&=&\frac{\epsilon^{abc}}{\sqrt{6}}\int[DX_3](-k_{2\perp}^-\psi^{(5)}(1,2,3))(u_{a\uparrow}^\dagger(1)u_{b\uparrow}^\dagger(2)d_{c\uparrow}^\dagger(3)
\nonumber\\
&&-u_{a\uparrow}^\dagger(1)d_{b\uparrow}^\dagger(2)u_{c\uparrow}^\dagger(3))\left|0\right\rangle\label{eq:LC3}\end{eqnarray}
\begin{eqnarray}\left|P\uparrow\right\rangle^{l_z=2}&=&\frac{\epsilon^{abc}}{\sqrt{6}}\int[DX_3]k_{1\perp}^+k_{3\perp}^+\psi^{(6)}(1,2,3)(u_{a\downarrow}^\dagger(1)d_{b\downarrow}^\dagger(2)u_{c\downarrow}^\dagger(3)
\nonumber\\
&&-u_{a\downarrow}^\dagger(1)u_{b\downarrow}^\dagger(2)d_{c\downarrow}^\dagger(3))\left|0\right\rangle\label{eq:LC4}
\end{eqnarray}
where $k_{i\perp}^{\pm}=k_i^x\pm ik_i^y$, while $u_{ai}^\dagger(1)$ means
the creation operator of an up-quark (same for down-quark) with
color $a$, spin $i$ and momentum $k_1$ etc, satisfying the light-cone
anti-commutation relation:
\begin{equation}\{u_{ai}(p),u_{bj}^\dagger(p')\}=2p^+(2\pi)^3\delta_{ab}\delta_{ij}\delta(p^+-p'^+)\delta^{(2)}(\vec{p}_\perp-\vec{p}_\perp ')\end{equation}
The integration measure is \footnote{There might be difference in
constant factors in the definition of integration measure by
different authors, which only affects the overall normalization.}:
\begin{eqnarray}\int[DX_3]&=&\sqrt{2}\frac{dx_1dx_2dx_3}{\sqrt{2x_12x_22x_3}}\frac{d^2\vec{k}_{1\perp} d^2\vec{k}_{2\perp}
d^2\vec{k}_{3\perp}}{(2\pi)^9}2\pi\delta(1-x_1-x_2-x_3)\times\nonumber\\
&&(2\pi)^2\delta^{(2)}(\vec{k}_{1\perp}+\vec{k}_{2\perp}+\vec{k}_{3\perp})\end{eqnarray}
The proton wavefunction amplitudes $\{\psi^{(1)}...\psi^{(6)}\}$ are
generally unknown functions. Although the expansion
\eqref{eq:LC1}$\sim$\eqref{eq:LC4} is generic,  the explicit form of
$\psi^{(i)}$ is model-dependent. In this work, we chose the form of
$\psi^{(i)}$ derived in Ref.~\cite{Pasquini_1} by starting from the
static solution of a constituent quark model \cite{CQM} (which works
well in predicting many electroweak properties of the baryons) and
applying a Melosh rotation to the solution to obtain non-zero $l_z$
components \cite{linking}. This choice of proton wavefunction is
used to predict the first moment of Sivers function, and turns out
to agree fairly well with the experimental measurements from HERMES
and COMPASS \cite{HERMES}\footnote{Ref. \cite{Pasquini_1} and Ref.
\cite{HERMES} defined their first moment of Sivers function with a
sign difference.}.

\section{Matrix elements between nucleon states}
\label{sec:mes}

Following \cite{Balitsky_Braun}, in order to obtain the twist-four distribution function $\tilde{Q}_D(x)$ we need to evaluate the matrix elements between state $\left|D(P)\right\rangle$ of the following operators:
\begin{eqnarray}
Q_A(b,z)&\equiv&:\bar{u}(b_1 z)t^az\!\!\!/\gamma_5u(b_2z)\bar{d}(b_3z)t^az\!\!\!/\gamma_5d(b_4z):\nonumber\\
Q_V(b,z)&\equiv&:\bar{u}(b_1 z)t^az\!\!\!/u(b_2z)\bar{d}(b_3z)t^az\!\!\!/d(b_4z):\label{eq:two_matrices}
\end{eqnarray}
where $z$ is a coordinate on light cone, and the parameters $b\equiv\{b_1,b_2,b_3,b_4\}$ characterize the light-cone separation between quark field operators.

When computing the matrix elements of $Q_{V,A}$ in
Eq.~\eqref{eq:two_matrices} we assume an incoherent impulse
approximation in which the incoming photon strikes only one of the
two nucleons (see, {\em e.g.} Ref. \cite{impulse_approx} for further
discussions regarding the impulse approximation); hence, matrix
elements of the operators \eqref{eq:two_matrices} can be related to
the same matrix elements taken between proton states (or
equivalently between neutron states, given isospin symmetry). Also,
since the quantities we compute do not depend on the proton spin, we
can take it to be +1/2 along the $z$-direction without loss of
generality.

Now, starting from the operators \eqref{eq:two_matrices}, we define two distribution functions $Q_\pm(x_\xi)$ via
\begin{equation}
\left\langle P(p)\uparrow\right|\{ Q_V(b,z)\pm Q_A(b,z)\}\left|P(p)\uparrow\right\rangle\equiv(p\cdot z)^2\int\prod_{k=1}^4d x_{\xi_k}\delta(\sum_i x_{\xi_i})e^{-i(p\cdot z)\sum_k b_k x_{\xi_k}}Q_\pm(x_\xi)\label{eq:Qpm}\end{equation}
with $x_\xi$ collectively representing $\{x_{\xi_1},x_{\xi_2},x_{\xi_3},x_{\xi_4}\}$,  the light-cone momentum fractions: $\xi_i^+=x_{\xi_i}p^+$.
Meanwhile $\left| P(p)\uparrow\right\rangle$ is the spin-up proton state with momentum $p$. Substituting \eqref{eq:LC1}$\sim$\eqref{eq:LC4} into \eqref{eq:Qpm} we are able
to express $Q_\pm(x_\xi)$ in terms of the proton wavefunction amplitudes. It is easy to observe that only diagonal terms, ({\em i.e.} terms with the same $l_z$ in initial and final states), could give non-vanishing contributions.
After a rather lengthy derivation with the aid of Eq.~\eqref{eq:four_fermion}, we obtain:
\begin{eqnarray}Q_{\pm}(x_\xi)&=&-\frac{32\pi^3}{3}\int\frac{d^2\vec{\xi}_{1\perp}}{(2\pi)^3}...\frac{d^2\vec{\xi}_{4\perp}}{(2\pi)^3}\theta(-x_{\xi_1})
\theta(x_{\xi_2})\theta(-x_{\xi_3})\theta(x_{\xi_4})\theta(1-x_{\xi_2}-x_{\xi_4})\times\nonumber\\
&&\delta^2(\vec{\xi}_{1\perp}+\vec{\xi}_{2\perp}+\vec{\xi}_{3\perp}+\vec{\xi}_{4\perp})\sum_{l_z}\psi^{\pm}_{l_z}(-\xi_1,-\xi_3,\xi_2,\xi_4)\label{eq:Qpmformula}\end{eqnarray}
where the explicit formulas of $\psi^{\pm}_{l_z}$ are given in Appendix \ref{sec:appB}.

The proton twist-four distribution function can now be expressed in
terms of the $Q_\pm$ (refer to Eq. (42) of Ref. \cite{Belitsky}
after some rearrangement):
\begin{eqnarray}
\tilde{Q}_p
(x_B)&\equiv&2\mathrm{Re}\int_{-1}^{1}\frac{\prod_{k=1}^4 dx_{\xi_k}}{x_{\xi_2}x_{\xi_3}(x_{\xi_2}+x_{\xi_3})}\delta(\sum_k x_{\xi_k})\{(x_{\xi_2}+x_{\xi_3})\delta(x_B+x_{\xi_1}+x_{\xi_2})-x_{\xi_3}\delta(x_B+x_{\xi_1})\nonumber\\
&&-x_{\xi_2}\delta(x_{\xi_4}-x_B)\}
[(1+P_{14}P_{23})Q_+(x_\xi)-(P_{12}+P_{34})Q_-(x_\xi)]\label{eq:Qpformula}
\end{eqnarray}
Here $P_{ij}$ is the permutation operator, e.g.
$P_{12}Q_+(x_{\xi_1},x_{\xi_2},x_{\xi_3},x_{\xi_4})=Q_+(x_{\xi_2},x_{\xi_1},x_{\xi_3},x_{\xi_4})$.
The deuteron twist-four distribution function $\tilde{Q}_D(x_B)$ can
be expressed in terms of $\tilde{Q}_p(x_B)$ through an incoherent
impulse approximation \cite{Impulse}, which says that a general
deuteron hadronic tensor can be related to the corresponding
hadronic tensors of proton and neutron by: \begin{equation} M_D
W^{\mu\nu}_D(p,q)\approx M_N
W^{\mu\nu}_p(\frac{p}{2},q)+M_NW^{\mu\nu}_n(\frac{p}{2},q)\label{eq:impulse}\end{equation}
where $M_N$ is the mass of nucleon. In the equation above each
hadronic tensor is multiplied by the particle's mass, because
following Eq. \eqref{eq:HT_matrix} the hadronic tensor we defined
has dimension -1. Now we can express both sides of Eq.
\eqref{eq:impulse} in terms of dimensionless structure functions
\{$F_i(x_B)$\}. Using isospin symmetry and the fact that
$\tilde{Q}(x_B)$ is proportional to $x_B^{-1} F_1^{ud}(x_B)$ (see
Eq. (34) of Ref. \cite{Belitsky}), we obtain \footnote{In Ref.
\cite{Belitsky}, the authors did not multiply their hadronic tensors
by particle mass in the impulse approximation formula, therefore the
corresponding relation they obtained is off by a factor 1/2; same
for the relation of quark distribution functions.}:
\begin{equation}
\frac{1}{2}\tilde{Q}_D(x_B/2)\approx
\tilde{Q}_p(x_B)+\tilde{Q}_n(x_B)\approx 2
\tilde{Q}_p(x_B)\end{equation}

Finally, following the logic of Ref.~\cite{Belitsky}, one can the derive the twist-four contribution to $R_1$:

\begin{equation}R_1^{HT}(x_B,Q^2)=\frac{1}{Q^2}\frac{\alpha_s\pi}{5(1-\frac{20}{9}\mathrm{sin}^2\theta_W)}\frac{x_B\tilde{Q}_D(x_B)}{u_D(x_B)+d_D(x_B)}\label{eq:central_eqn}\end{equation}
with $q_D(x_B)$ being the parton distribution function for quark of flavor $q$ in the deuteron
\begin{equation}
\left\langle
D(P)\right|\bar{q}(z)z\!\!\!/q(-z)\left|D(P)\right\rangle=2(P\cdot
z)\int_{-1}^{1}dx e^{2i(P\cdot z)x}q_D(x)\label{eq:qdf}
\end{equation}
Note that we neglect the logarithmic $Q^2$-dependence of the
structure functions in this analysis. We can express $q_D$ in terms
of PDF of the proton and neutron again by the impulse approximation
\eqref{eq:impulse}, but now comparing the structure function
$F_2(x_B)$ on both sides, which is proportional to $x_B^{-1}q(x_B)$.
The result is:
\begin{equation}
q_D(x_B/2)\approx q_p(x_B)+q_n(x_B )\end{equation} where $q_p(x)$
and $q_n(x)$ are defined as in Eq.~\eqref{eq:qdf} but for
proton/neutron states. Furthermore, neglecting  CSV effects we have:
\begin{equation}
u_n(x_B)=d_p(x_B), d_n(x_B)=u_p(x_B)
\end{equation}
Therefore, it is sufficient to just calculate $u_p(x_B)$ and
$d_p(x_B)$ using the proton light-cone wavefunction
\eqref{eq:LC1}$\sim$\eqref{eq:LC4}. Using \eqref{eq:two_fermion1}
and \eqref{eq:two_fermion2} , we can compute the quark PDFs of the
(spin-up) nucleons by calculating the matrix element on LHS of Eq.
\eqref{eq:qdf} with nucleon states, and compare it with the form on
RHS to extract the PDFs. Same with the twist-four distribution
functions, only terms diagonal to $l_z$ survive, so we can separate
the result into components of different $l_z$ as the following:
\begin{eqnarray}u_p(x_B)+d_p(x_B)&=&d_n(x_B)+u_n(x_B)\nonumber\\
&=&\frac{1}{(2\pi)^6}\int_0^1 dx_1\int
d^2\vec{k}_{1\perp}d^2\vec{q}_{\perp}\Theta(1-x_B-x_1)\sum_{l_z}A^{l_z}(q,1,2)\label{eq:qdfformula}\end{eqnarray} where the functions $A^{l_z}(q,1,2)$ are
given in Appendix \ref{sec:appB}.

We now proceed to show that a partial cancelation occurs
between contributions of $l_z=+1$ and $l_z=-1$. For this purpose, we
combine \eqref{eq:Qpmformula} and \eqref{eq:Qpformula}, together
with the fact that
$\psi^{\pm}_{l_z}(q,l,q',l')^*=\psi^{\pm}_{l_z}(q',l',q,l)$, to
simplify the expression of $\tilde{Q}_p(x_B)$ as:
\begin{equation}
\tilde{Q}_p(x_B)=\tilde{Q}^+_p(x_B)+\tilde{Q}^-_p(x_B)
\end{equation}
where
\begin{eqnarray}
\tilde{Q}^+_p(x_B)&=&\frac{64\pi^3}{3}\int_0^1\prod_{i=1}^4dx_{\xi_i}\delta(x_{\xi_1}-x_{\xi_2}+x_{\xi_3}-x_{\xi_4})\theta(1-x_{\xi_2}-x_{\xi_4})\{\frac{\delta(x_B-x_{\xi_1}+x_{\xi_2})}{x_{\xi_2}x_{\xi_3}}\nonumber\\
&&+\frac{\delta(x_B-x_{\xi_1})}{x_{\xi_2}(x_{\xi_2}-x_{\xi_3})}-\frac{\delta(x_B-x_{\xi_4})}{x_{\xi_3}(x_{\xi_2}-x_{\xi_3})}+\frac{\delta(x_B+x_{\xi_3}-x_{\xi_4})}{x_{\xi_1}x_{\xi_4}}\}\times\nonumber\\
&&\int\prod_{i=1}^4\frac{d^2\vec{\xi}_{i\perp}}{(2\pi)^3}\delta^2(\vec{\xi}_{\perp1}-\vec{\xi}_{\perp2}+\vec{\xi}_{\perp3}-\vec{\xi}_{\perp4})\sum_{l_z}\mathrm{Re}\psi_{l_z}^+(\xi_1,\xi_3,\xi_2,\xi_4)\label{eq:Qpplus}
\end{eqnarray}
\begin{eqnarray}
\tilde{Q}^-_p(x_B)&=&\frac{64\pi^3}{3}\int_0^1\prod_{i=1}^4dx_{\xi_i}\delta(x_{\xi_1}-x_{\xi_2}-x_{\xi_3}+x_{\xi_4})\theta(1-x_{\xi_2}-x_{\xi_3})\{\frac{\delta(x_B+x_{\xi_1}-x_{\xi_2})}{x_{\xi_2}x_{\xi_3}}\nonumber\\
&&-\frac{\delta(x_B-x_{\xi_4})}{x_{\xi_3}(x_{\xi_2}+x_{\xi_3})}-\frac{\delta(x_B-x_{\xi_1})}{x_{\xi_2}(x_{\xi_2}+x_{\xi_3})}+\frac{\delta(x_B-x_{\xi_1}+x_{\xi_2})}{x_{\xi_2}x_{\xi_3}}\}\times\nonumber\\
&&\int\prod_{i=1}^4\frac{d^2\vec{\xi}_{i\perp}}{(2\pi)^3}\delta^2(\vec{\xi}_{\perp1}-\vec{\xi}_{\perp2}-\vec{\xi}_{\perp3}+\vec{\xi}_{\perp4})\sum_{l_z}\mathrm{Re}\psi_{l_z}^-(\xi_2,\xi_3,\xi_1,\xi_4)\label{eq:Qpminus}
\end{eqnarray}

First we qualitatively analyze the contribution from each
$l_z$-component to $\tilde{Q}^{\pm}_p(x_B)$. This can be done by
simply referring to
Eqs.~\eqref{eq:lz0_begin}$\sim$\eqref{eq:lz2_end} of the Appendix B.
The result is summarized in Table \ref{tab:contribution}. We observe
that the $l_z=+1$ (-1) piece contributes mainly to $\tilde{Q}^-_p$
$(\tilde{Q}^+_p)$. Also notice that we do not include the $l_z=2$
component as its effect is tiny.

\begin{table}
\caption{\label{tab:contribution}The contributions from different
$l_z$-components to $\tilde{Q}^{\pm}_p(x_B)$. The $l_z$=0,+1
components contribute mostly to $\tilde{Q}_p^-$ (\lq\lq dominant")
and less so to $\tilde{Q}_p^+$ (\lq\lq subdominant"), while the
$l_z$=-1 component  contributes only to $\tilde{Q}_p^+$.}
\begin{ruledtabular}
\begin{tabular}{ccc}
$l_z$&Contribution to $\tilde{Q}^+_p(x_B)$&Contribution to $\tilde{Q}^-_p(x_B)$\\
\hline
0&subdominant&dominant\\
+1&subdominant&dominant\\
-1&all&zero\\
\end{tabular}
\end{ruledtabular}
\end{table}

\begin{figure}
\includegraphics[scale=0.35]{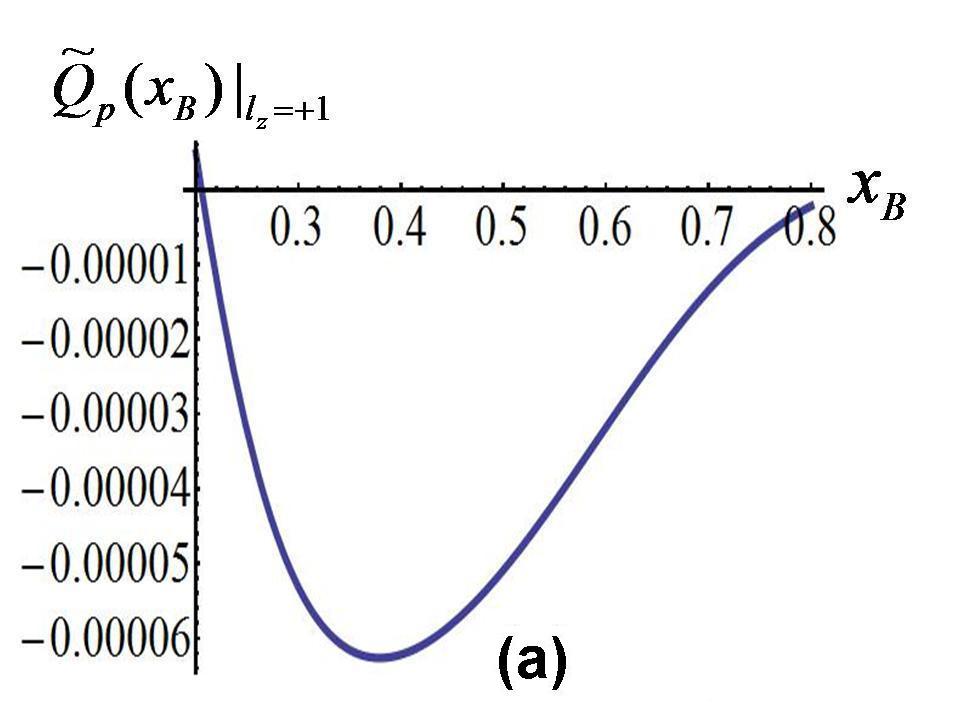}\includegraphics[scale=0.35]{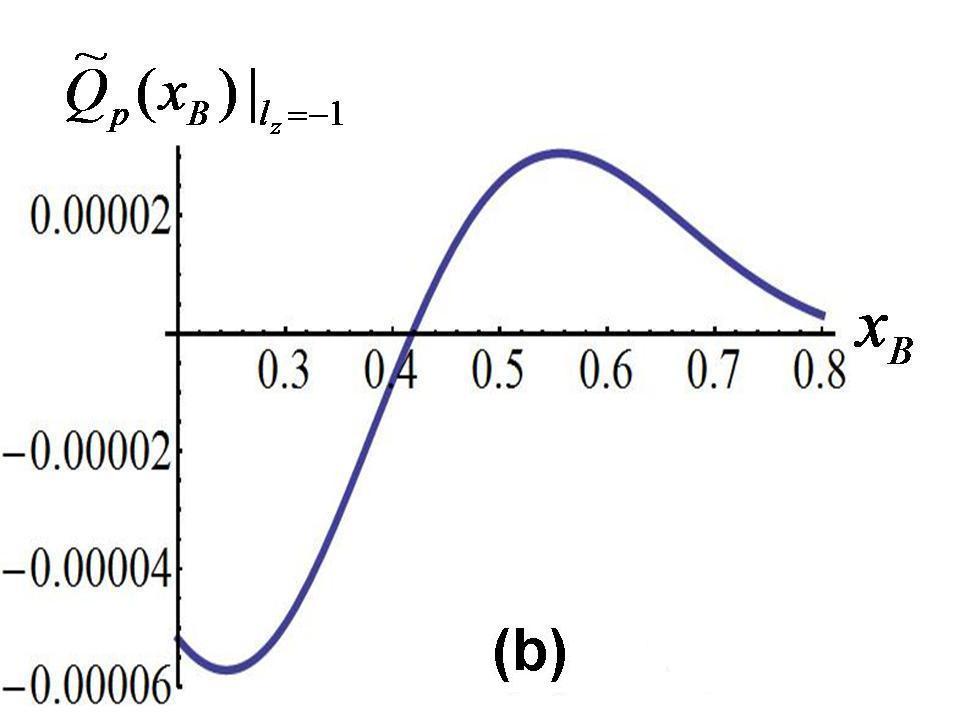}\\
\includegraphics[scale=0.35]{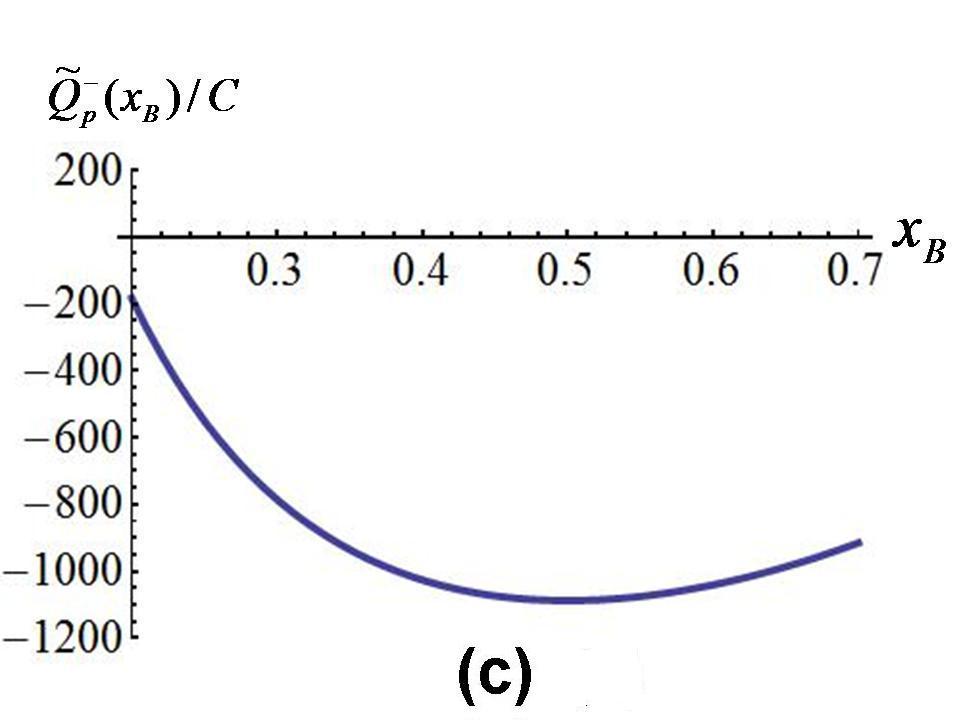}\includegraphics[scale=0.35]{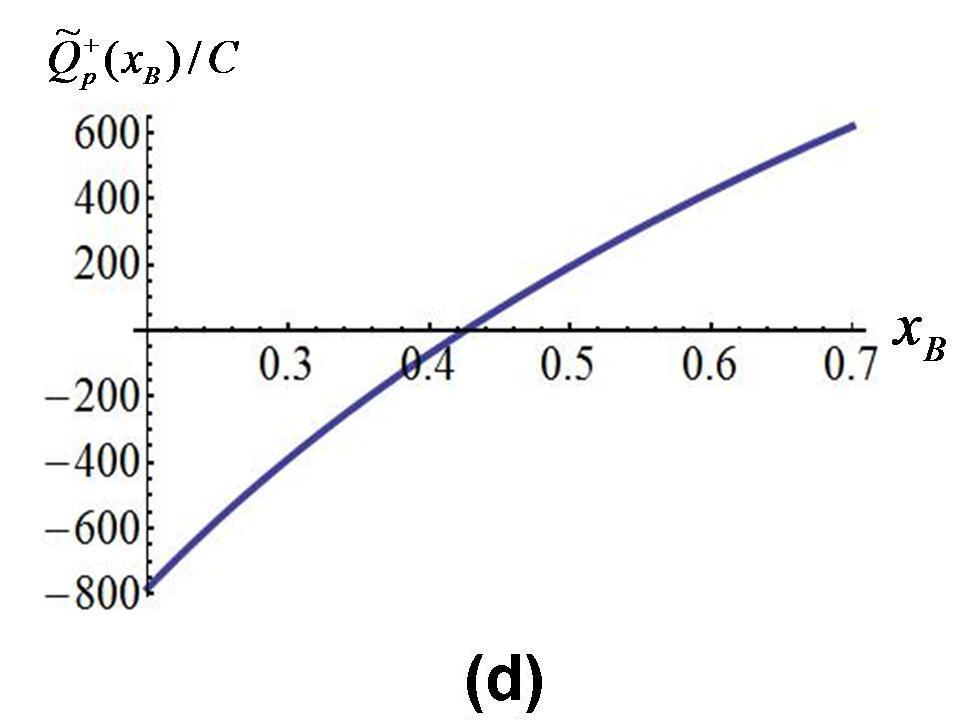}
\caption{\label{fig:Qpm}(Color online) Top panels: full results for
$l_z\pm 1$ contributions to $\tilde{Q}_p(x_B)$. Bottom panels:
behavior of  $\tilde{Q}^{\mp}_p(x_B)$ ignoring the details of
nucleon wavefunction amplitudes. The constant C is defined in Eq.
\eqref{eq:approximation}.}
\end{figure}

Next we study the behavior of different contributions to
$\tilde{Q}^{\pm}_p(x_B)$ with respect to $x_B$, showing that those
associated with the $l_z\pm 1$ components largely cancel. The
individual contributions from the latter are shown in the top two
panels of Fig.~\ref{fig:Qpm}. We observe that the $l_z=-1$
contribution, which contributes only to $\tilde{Q}^{+}_p(x_B)$
changes sign at $x_B\approx 0.4$, whereas the $l_z=+1$ contribution
does not. Consequently, the two contributions will cancel against
each other for $x_B\gsim 0.4$. While the cancellation is not exact,
it becomes more effective at larger values of $x_B$, a region that
is weighted most strongly in
 $R_1^{HT}$ by the factor of $x_B$ in the numerator of Eq.~(\ref{eq:central_eqn}) and the corresponding presence of $u_D(x_B)+d_D(x_B)$ in the denominator.

We also note that this sign change and cancellation appears to be
rather generic. To see why, let us naively take:
\begin{equation}
\int\prod_{i=1}^4\frac{d^2\vec{\xi}_{i\perp}}{(2\pi)^3}\delta^2(...)\mathrm{Re}\psi_{l_z}^{\pm}\approx\mathrm{constant}\equiv
C\label{eq:approximation}
\end{equation}
assuming the function above is well-behaved with respect to
\{$x_{\xi_i}$\}. This approximation simply means that we do not care
about the details of the proton wavefunction amplitudes. Under this
approximation, the numerical integration \eqref{eq:Qpplus} and
\eqref{eq:Qpminus} can be performed quite trivially, and the result
is shown in the lower two panels of Fig \ref{fig:Qpm}. In this case,
we show $\tilde{Q}^{\pm}_p(x_B)$ as the $l_z=\pm 1$ components
contribute primarily to one or the other of these two quantities
(see Table \ref{tab:contribution}). Although the the assumption in
Eq.~(\ref{eq:approximation}) breaks down at large and small $x_B$,
one can see that a sign change of $\tilde{Q}^+_p(x_B)$ from negative
to positive occurs near $x_B=0.4$, implying that
$\tilde{Q}^+_p(x_B)$ and $\tilde{Q}^-_p(x_B)$ will have different
signs for $x_B \gsim 0.4$. Therefore, according to Table
\ref{tab:contribution}, the contribution to $\tilde{Q}_p(x_B)$ from
$l_z=1$ and $l_z=-1$ should partially cancel  other for $x_B\gsim
0.4$. Furthermore, since the argument above does not depend on the
details of the nucleon wavefunction (as long as it is well-behaved),
this feature of partial cancelation should be generic.


\section{Numerical results and discussion}

Eqs.~\eqref{eq:Qpplus} and \eqref{eq:Qpminus} are our starting point
for the numerical evaluation of $\tilde{Q}_p(x_B)$, which involves
an eight-fold integration. To perform this integration, we adopt the
Monte Carlo numerical integration called Divonne contained in the
CUBA Library, which is an algorithm package designed for
multi-dimensional numerical integration \cite{Cuba}. For each $l_z$
component, we evaluate the value of $\tilde{Q}_p(x_B)$ at a series
of discrete $\{x_{B,i}\}$, and then link them together using a
best-fit line. Also, we take $\alpha_s=0.5$ at 1GeV following the
renormalization group (RG) prediction of the running coupling
constant at 4-loop order together with a 3-loop threshold
matching, with the quark thresholds taken to be
$M_c=1.5$ GeV and $M_b=4.7$ GeV respectively \cite{Running}.

\begin{figure}
\includegraphics[scale=0.35]{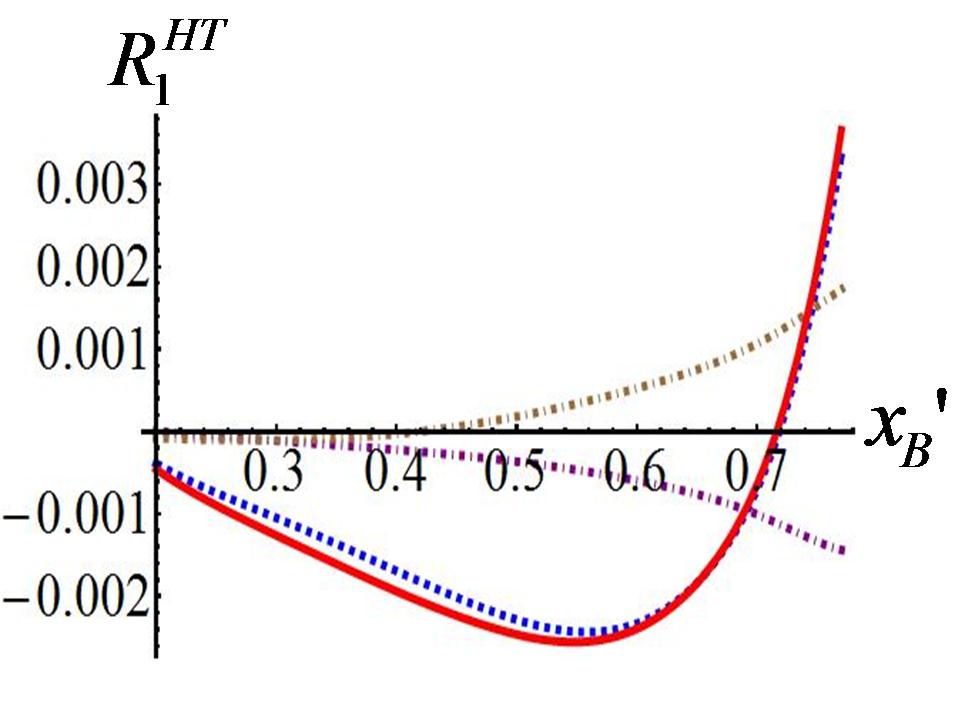}\caption{\label{fig:HT_correction}(Color online)The Twist-4 correction to $R_1$ at $Q^2=4\mathrm{GeV}^2$. The blue dashed curve
shows the $l_z=0$ contribution; purple dot-dashed curve shows the
$l_z=1$ contribution; brown dot-dashed curve shows the $l_z=-1$
contribution; the red solid curve is the sum of all. $l_z=2$
contribution is negligible and therefore not included.}
\end{figure}

Our main result is shown in Fig. \ref{fig:HT_correction}, which gives $R_1^{HT}$ versus $x_B'\equiv 2x_B$ at $Q^2=4\mathrm{GeV}^2$.
First, let us compare this outcome with that of Refs. \cite{Sonny} and
\cite{Belitsky}. It turns out that all three calculations predict
similar curve shape for $R_1^{HT}$, only with slightly different
positions of peak and zero-point. Concerning the magnitude, our work
predicts a maximum absolute value
$|R_1^{HT}|\approx2.6\times10^{-3}$ between $0.2<x_B'<0.7$, which is smallest in magnitude among all the three predictions, and is about a half of
the size to that of Ref. \cite{Belitsky}. This is understandable
because the authors include a 3-quark+1-gluon Fock-space component whose contribution
is comparable in magnitude to that of the pure 3-quark state. Nonetheless, all three
calculations suggest that $|R_1^{HT}|$ lies below that of the expected SoLID precision.

Next we study the OAM-dependence in detail. To that end, we first
introduce some nomenclature: in the following, we will use the
notation $(|l_z|\otimes |l_{z'}|)$, which denotes a generic matrix
element taking between two hadronic states, of which one of them has
absolute value of quark OAM in z-direction being $|l_z|$ and the
other being $|l_{z'}|$.

\begin{figure}
\includegraphics[scale=0.3]{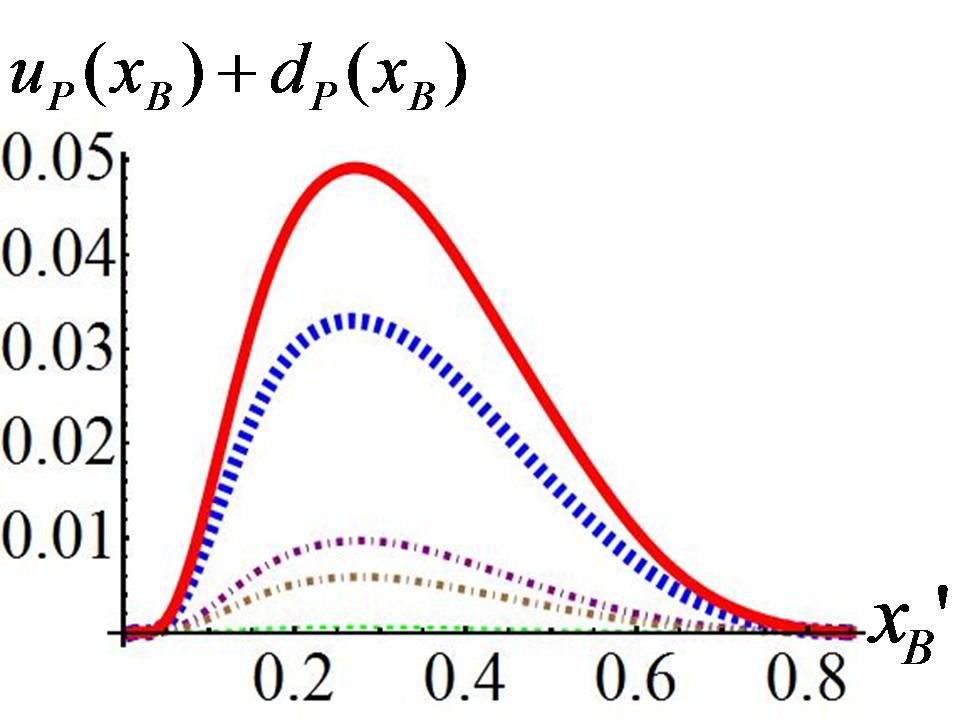}\caption{\label{fig:QDF}(color online)The unnormalized QDF of spin-up proton, splitted into contributions from different $l_z$ components. Blue thick-dashed curve shows contribution
from $l_z=0$ component; purple dot-dashed curve shows contribution
from $l_z=1$ component; brown dot-dashed curve shows contribution
from $l_z=-1$ component; green thin-dashed curve shows contribution
from $l_z=2$ component; red solid curve is the sum of all
contributions.}
\end{figure}

From our arguments at the end of Section \ref{sec:mes}, we expect
that although $l_z=\pm1$ individually contribute a significant
amount to $\tilde{Q}_p(x_B)$, they should largely cancel against
each other for $x_B>0.4$, making the total $(1\otimes1)$
contribution rather small, and therefore leaving the $(0\otimes0)$
contribution as the dominant piece. This expectation is born out by
the curves in Fig. \ref{fig:HT_correction}. The purple dot-dashed
curve and brown dot-dashed curve curves give the individual $(l_z =
1)\otimes ( l_z= 1)$ and $(l_z = -1)\otimes ( l_z= -1)$
contributions, respectively , which exhibit the expected
cancellation for $x_B'> 0.4$. The blue dashed curve and red solid
curve give the $(0\otimes 0)$ and total contributions, respectively.
It is clear that the former dominates the total. This $(0\otimes 0)$
dominance is a rather unique feature of the particular twist-four
contribution of interest here, and one that
 is not shared by other diagonal matrix elements. For example,
if one calculate proton quark PDFs (leading twist) using the same set of wavefunctions, the
$(0\otimes0)$ and $(1\otimes1)$ contributions are comparable; moreover,
since they have the same sign, the two $|l_z|=1$ pieces do not
cancel each other (see Fig.\ref{fig:QDF}).

On the other hand, we also note that there are hadronic matrix
elements that depend crucially on the existence of non-zero quark
OAM in light cone quantization. In particular, in
Ref.~\cite{Pasquini_1}, the authors studied the Sivers function
\cite{Sivers} and Boer-Mulders function \cite{Boer_Mulders}, which
are examples of transverse momentum dependent parton distribution
functions (TMDs), appearing in semi-inclusive deep inelastic
scattering. Importantly, both distribution functions depend on
off-diagonal matrix elements of $l_z$: the Sivers function is
sensitive to $(0\otimes1)$ while Boer-Mulders function is sensitive
to both $(0\otimes1)$ and $(1\otimes2)$. Simply speaking, the
existence of non-zeo quark OAM is responsible for the non-vanishing
values of the Sivers and Boer-Mulders functions. Combining this
observation with our analysis of the HT matrix element, we conclude
that the twist-four correction to $e$D PVDIS is
 essentially  transparent to the parton angular
momentum dynamics that generate the Sivers and Boer-Mulders
functions.

It is also interesting to study the impact of sea-parton dynamics on
the behavior of the HT matrix element. To that end, we performed a
qualitative analysis of the contribution made by the Fock space
component containing  3 quarks + 1 gluon, using the general form
suggested in Ref. \cite{3q1g} that includes non-zero gluon OAM. The
authors of Ref.~\cite{Belitsky} computed the contribution of the
3q+1g state with $l_z=0$ , which turns out to have a  similar shape
to that of the $l_z=0$  3q-state contribution. To our knowledge,
however, there exist no explicit functional forms for the 3q+1g
nucleon wavefunction with non-zero parton OAM. Consequently, our
analysis is purely analytic at this point. We observe that, in
contrast to the 3q state contribution, the matrix element of 3q+1g
state for a fixed $l_z$ can contribute significantly to both
$\tilde{Q}_p^{\pm}(x_B)$ simultaneously; therefore there is no
obvious correlation between $l_z$ and $\tilde{Q}_p^{\pm}(x_B)$ and
hence no obvious pattern of partial cancelation. In Table
\ref{tab:table} we summarize the importance of different
$(|l_z|\otimes|l_z'|)$ contributions to various distribution
functions, considering only the contributions of 3q states.

Combining observations, we may draw the following
conclusion: if a future $e$D PVDIS measurement yields a sufficiently precise
determination of $R_1^{HT}$ as a function of $x_B'$, one can
compare the experimental curve with our current theoretical
prediction. A significant deviation from the predicted curve ({\em e.g.}, the peak and zero-point are shifted
by a considerable amount), could signal the importance of parton angular momentum dynamics
beyond those responsible for the Sivers, Boer-Mulders, and spin-independent parton distribution functions.

\begin{table}
\caption{\label{tab:table}The dependence on different quark light-cone OAM
components of various distribution functions.}
\begin{ruledtabular}
\begin{tabular}{ccc}
Distribution Functions&Dominant Contribution(s)&Subdominant Contribution(s)\\
\hline
Quark Distribution Functions&(0$\times$0), (1$\otimes$1)&(2$\otimes$2)\\
PVDIS Twist-Four Correction&(0$\otimes$0)&(1$\otimes$1), (2$\otimes$2)\\
Sivers Function&(0$\otimes$1)&(1$\otimes$2)\\
Boer-Mulders Function&(0$\otimes$1), (1$\otimes$2)&---
\end{tabular}
\end{ruledtabular}
\end{table}

\section{Summary}

The next generation of parity-violating electron scattering
experiments are poised to probe both possible BSM physics as well as
novel features of hadron and nuclear structure. In this work, we
have studied one particular hadronic effect, namely, the twist-four
contribution to $\tilde{a}_1$, the $y$-independent term in the PV
asymmetry. Using a set of proton light-cone wavefunctions with
non-zero quark orbital angular momentum, we evaluated the twist-four
contribution as a function of $x_B$, identifying the  contributions
from different OAM-components. Our total for the correction
$R_1^{HT}$ is similar in both shape and magnitude to those obtained
in previous works, indicating that  higher-precision than expected
with the SoLID experiment would be needed to discern this HT effect.
An effort to achieve such precision may be worthwhile, because
$R_1^{HT}$ appears to be rather unique, in the sense that it is not
significantly affected by the parton angular momentum physics
responsible for the existence of some other DIS observables such as
the Sivers and Boer-Mulders functions. Thus, by combining the
results of a more precise measurement of the asymmetry with
measurements of other distribution functions, it is possible to
probe complementary aspects of parton angular momentum
 and, perhaps, shed new light
on the role of angular momentum in the structure of the nucleon.

\section{Acknowledgements}

We would like to thank A. Belitksy for useful discussions and F. Yuan and B. Pasquini for pointing us to
explicit examples of the nucleon wavefunction amplitudes. We are also grateful to F. Yuan for
a careful reading of the manuscript. This work
is supported in part by DOE Contract DE-FG02-08ER41531 and by the
Wisconsin Alumni Research Foundation.

\appendix

\section{Matrix Elements of Two and Four-Fermion Operators}

In this section we present matrix elements of two-fermion operators
($u^\dagger u$ and $d^\dagger d$) and four-fermion operators
($u^\dagger ud^\dagger d$) between nucleon states. For this purpose
let us consider two arbitrary components of proton light-cone
wavefunction defined as the following:
\begin{eqnarray}\left|\psi_\alpha\right\rangle&\equiv&\frac{\epsilon^{abc}}{\sqrt{6}}\int [DX_3] \psi_\alpha (1,2,3)u^{\dagger}_{a\lambda_1}(1)
u^{\dagger}_{b\lambda_2}(2)d^{\dagger}_{c\lambda_3}(3)\left|0\right\rangle\nonumber\\
\left|\psi_\beta\right\rangle&\equiv&\frac{\epsilon^{abc}}{\sqrt{6}}\int
[DX_3] \psi_\beta (1,2,3)u^{\dagger}_{a\lambda_1'}(1)
u^{\dagger}_{b\lambda_2'}(2)d^{\dagger}_{c\lambda_3'}(3)\left|0\right\rangle\label{eq:states}\end{eqnarray}
It is straightforward to work out the matrix elements of the
four-fermion operator between these two states (the symbol ``1"
denotes the four momentum $k_1=(x_1 p^+,\vec{k}_{1\perp})$ which is
given by $x_1=1-x_q-x_l=1-x_q'-x_l'$ and
$\vec{k}_{1\perp}=-\vec{q}_{\perp}-\vec{l}_{\perp}=-\vec{q}'_{\perp}-\vec{l}'_{\perp}$.):
\begin{eqnarray}&&\left\langle\psi_\alpha\right|u^{\dagger}_{i\rho}(q)u_{i'\rho'}(q')d^{\dagger}_{j\lambda}(l)d_{j'\lambda'}(l')\left|\psi_\beta\right\rangle
=\frac{32\pi^3}{3}(\delta_{ii'}\delta_{jj'}-\delta_{ij'}\delta_{i'j})\delta_{\lambda_3\lambda}\delta_{\lambda_3'\lambda'}\sqrt{x_qx_lx_q'x_l'}\nonumber\\
&&\delta(x_q+x_l-x_q'-x_l')\delta^2(\vec{q}_\perp+\vec{l}_\perp-\vec{q'}_\perp-\vec{l'}_\perp)\int
dx_1d^2\vec{k_1}_\perp\delta(1-x_1-x_q-x_l)\delta^2(\vec{k_1}_\perp+\vec{q}_\perp+\vec{l}_\perp)\nonumber\\
&&(\delta_{\lambda_1\rho}\delta_{\lambda_2\lambda_2'}
\delta_{\rho'\lambda_1'}\psi_\alpha^*(q,1,l)\psi_\beta(q',1,l')+\delta_{\lambda_1\lambda_2'}\delta_{\lambda_2\rho}\delta_{\rho'\lambda_1'}\psi_\alpha^*(1,q,l)\psi_\beta(q',1,l')
\nonumber\\&&+\delta_{\lambda_1\rho}\delta_{\lambda_2\lambda_1'}\delta_{\rho'\lambda_2'}\psi_\alpha^*(q,1,l)\psi_\beta(1,q',l')+\delta_{\lambda\lambda_1'}\delta_{\lambda_2\rho}\delta_{\rho'\lambda_2'}\psi_\alpha^*(1,q,l)\psi_\beta(1,q',l'))\label{eq:four_fermion}
\end{eqnarray}
and those for two-fermion operators:
\begin{eqnarray}\left\langle\psi_\alpha\right|d^{\dagger}_{j\lambda}(l)d_{j'\lambda'}(l')\left|\psi_\beta\right\rangle
&=&\frac{4}{3}x_l
\delta(x_l-x_{l}')\delta^2(\vec{l}_\perp-\vec{l}_\perp')\delta_{\lambda_3\lambda}\delta_{\lambda_3'\lambda'}\delta_{jj'}
\int dx_1
dx_2d^2\vec{k}_{1\perp}d^2\vec{k}_{2\perp}\nonumber\\
&&\delta(1-x_1-x_2-x_l)
\delta^2(\vec{k}_{1\perp}+\vec{k}_{2\perp}+\vec{l}_{\perp})(\delta_{\lambda_1\lambda_1'}\delta_{\lambda_2\lambda_2'}\psi_\beta(1,2,l)
\nonumber\\
&&+\delta_{\lambda_1\lambda_2'}\delta_{\lambda_2\lambda_1'}\psi_\beta(2,1,l))\psi_\alpha^*(1,2,l)\label{eq:two_fermion1}\end{eqnarray}
\begin{eqnarray}\left\langle\psi_\alpha\right|u^{\dagger}_{j\lambda}(l)u_{j'\lambda'}(l')\left|\psi_\beta\right\rangle
&=&\frac{4}{3}x_l
\delta(x_l-x_{l}')\delta^2(\vec{l}_\perp-\vec{l}_\perp')\delta_{\lambda_3\lambda_3'}\delta_{jj'}
\int dx_1
dx_2d^2\vec{k}_{1\perp}d^2\vec{k}_{2\perp}\nonumber\\
&&\delta(1-x_1-x_2-x_l)
\delta^2(\vec{k}_{1\perp}+\vec{k}_{2\perp}+\vec{l}_{\perp})(\delta_{\lambda_1\lambda}\delta_{\lambda_2\lambda_2'}\delta_{\lambda'\lambda_1'}
\psi_\alpha^*(l,1,2)\psi_\beta(l,1,2)\nonumber\\
&&+\delta_{\lambda_1\lambda_2'}\delta_{\lambda_2\lambda}\delta_{\lambda'\lambda_1'}\psi_\alpha^*(1,l,2)
\psi_\beta(l,1,2)+\delta_{\lambda_1\lambda}\delta_{\lambda_2\lambda_1'}\delta_{\lambda'\lambda_2'}\psi^*_\alpha(l,1,2)\psi_\beta(1,l,2)\nonumber\\
&&+
\delta_{\lambda_1\lambda_1'}\delta_{\lambda_2\lambda}\delta_{\lambda'\lambda_2'}\psi_\alpha^*(1,l,2)\psi_\beta(1,l,2))\label{eq:two_fermion2}\end{eqnarray}

\section{Complete formulae for quark PDFs and $\tilde{Q}_p(x_B)$ in terms of proton wavefunction amplitudes}
\label{sec:appB}

In this section we present explicit expressions needed to compute the quark PDFs and the twist-four distribution function.

The distribution functions $\tilde{Q}^{\pm}_p(x_B)$ in Eqs.~\eqref{eq:Qpplus} and \eqref{eq:Qpminus} are expressed in terms of $\psi^{\pm}_{l_z}(q,l,q',l')$,
which have the following expressions:

\begin{eqnarray}\psi^+_{l_z=0}(q,l,q',l')&=&2\psi^{(1,2)*}(q,1,l)\psi^{(1,2)}(q',1,l')\label{eq:lz0_begin}\\
\psi^-_{l_z=0}(q,l,q',l')&=&2\{\psi^{(1,2)*}(1,q,l)\psi^{(1,2)}(1,q',l')+\psi^{(1,2)*}(q,l,1)\psi^{(1,2)}(q',l',1)\nonumber\\
&&+\psi^{(1,2)*}(1,l,q)\psi^{(1,2)}(q',l',1)+\psi^{(1,2)*}(q,l,1)\psi^{(1,2)}(1,l',q')\nonumber\\
&&+\psi^{(1,2)*}(1,l,q)\psi^{(1,2)}(1,l',q')\}\\
\psi^+_{l_z=1}(q,l,q',l')&=&2\psi^{(3,4)*}(1,q,l)\psi^{(3,4)}(1,q',l')\label{eq:lz1_begin}\\
\psi^-_{l_z=1}(q,l,q',l')&=&2\{\psi^{(3,4)*}(q,1,l)\psi^{(3,4)}(q',1,l')+\psi^{(3,4)*}(l,q,1)\psi^{(3,4)}(l',q',1)\nonumber\\
&&+\psi^{(3,4)*}(l,1,q)\psi^{(3,4)}(l',q',1)+\psi^{(3,4)*}(l,q,1)\psi^{(3,4)}(l',1,q')\nonumber\\
&&+\psi^{(3,4)*}(l,1,q)\psi^{(3,4)}(l',1,q')\}\\
\psi^+_{l_z=-1}(q,l,q',l')&=&2\{\psi^{(5,5)*}(q,1,l)\psi^{(5,5)}(q',1,l')+\psi^{(5,5)*}(1,q,l)\psi^{(5,5)}(q',1,l')\nonumber\\
&&+\psi^{(5,5)*}(q,1,l)\psi^{(5,5)}(1,q',l')+\psi^{(5,5)*}(1,q,l)\psi^{(5,5)}(1,q',l')\}\\
\psi^-_{l_z=-1}(q,l,q',l')&=&0\label{eq:lz1_end}\\
\psi^+_{l_z=2}(q,l,q',l')&=&2\{\psi^{(6,6)*}(q,1,l)\psi^{(6,6)}(q',1,l')+\psi^{(6,6)*}(1,q,l)\psi^{(6,6)}(q',1,l')\nonumber\\
&&+\psi^{(6,6)*}(q,1,l)\psi^{(6,6)}(1,q',l')+\psi^{(6,6)*}(1,q,l)\psi^{(6,6)}(1,q',l')\}\\
\psi^-_{l_z=2}(q,l,q',l')&=&0\label{eq:lz2_end}\end{eqnarray} The
definitions of $\psi^{(i,j)}$ are the following:

\begin{eqnarray}\psi^{(1,2)}(1,2,3)&=&\psi^{(1)}(1,2,3)+i(k_1^x k_2^y-k_1^y k_2^x)\psi^{(2)}(1,2,3)\nonumber\\
\psi^{(3,4)}(1,2,3)&=&k_{1\perp}^+\psi^{(3)}(1,2,3)+k_{2\perp}^+\psi^{(4)}(1,2,3)\nonumber\\
\psi^{(5,5)}(1,2,3)&=&-k_{2\perp}^-\psi^{(5)}(1,2,3)+k_{3\perp}^-\psi^{(5)}(1,3,2)\nonumber\\
\psi^{(6,6)}(1,2,3)&=&k_{1\perp}^+(k_{2\perp}^+\psi^{(6)}(1,3,2)-k_{3\perp}^+\psi^{(6)}(1,2,3))\end{eqnarray}

On the other hand, the quark distribution functions in \eqref{eq:qdfformula} are given in terms of $A^{l_z}(q,1,2)$, which look like the
following:

\begin{eqnarray}A^{l_z=0}(q,1,2)&=&\psi^{(1,2)*}(q,1,2)\psi^{(1,2)}(q,1,2)+2\psi^{(1,2)*}(1,q,2)\psi^{(1,2)}(1,q,2)\nonumber\\
&&+\psi^{(1,2)*}(q,2,1)\psi^{(1,2)}(q,2,1)+\psi^{(1,2)*}(1,2,q)\psi^{(1,2)}(q,2,1)\nonumber\\
&&+\psi^{(1,2)*}(q,2,1)\psi^{(1,2)}(1,2,q)+2\psi^{(1,2)*}(1,2,q)\psi^{(1,2)}(1,2,q)\nonumber\\
&&+\psi^{(1,2)*}(1,q,2)\psi^{(1,2)}(2,q,1)\end{eqnarray}
\begin{eqnarray}A^{l_z=1}(q,1,2)&=&2\psi^{(3,4)*}(q,1,2)\psi^{(3,4)}(q,1,2)+\psi^{(3,4)*}(1,q,2)\psi^{(3,4)}(1,q,2)\nonumber\\
&&+\psi^{(3,4)*}(2,q,1)\psi^{(3,4)}(2,q,1)+\psi^{(3,4)*}(2,1,q)\psi^{(3,4)}(2,q,1)\nonumber\\
&&+\psi^{(3,4)*}(2,q,1)\psi^{(3,4)}(2,1,q)+\psi^{(3,4)*}(2,1,q)\psi^{(3,4)}(2,1,q)\nonumber\\
&&+\psi^{(3,4)*}(1,2,q)\psi^{(3,4)}(1,2,q)+\psi^{(3,4)*}(q,1,2)\psi^{(3,4)}(q,2,1)\end{eqnarray}
\begin{eqnarray}A^{l_z=-1}(q,1,2)&=&\psi^{(5,5)*}(q,1,2)\psi^{(5,5)}(q,1,2)+\psi^{(5,5)*}(1,q,2)\psi^{(5,5)}(q,1,2)\nonumber\\
&&+\psi^{(5,5)*}(q,1,2)\psi^{(5,5)}(1,q,2)+\psi^{(5,5)*}(1,q,2)\psi^{(5,5)}(1,q,2)\nonumber\\
&&+\psi^{(5,5)*}(1,2,q)\psi^{(5,5)}(1,2,q)+\psi^{(5,5)*}(1,2,q)\psi^{(5,5)}(2,1,q)\end{eqnarray}
\begin{eqnarray}A^{l_z=2}(q,1,2)&=&\psi^{(6,6)*}(q,1,2)\psi^{(6,6)}(q,1,2)+\psi^{(6,6)*}(1,q,2)\psi^{(6,6)}(q,1,2)\nonumber\\
&&+\psi^{(6,6)*}(q,1,2)\psi^{(6,6)}(1,q,2)+\psi^{(6,6)*}(1,q,2)\psi^{(6,6)}(1,q,2)\nonumber\\
&&+\psi^{(6,6)*}(1,2,q)\psi^{(6,6)}(1,2,q)+\psi^{(6,6)*}(1,2,q)\psi^{(6,6)}(2,1,q)\end{eqnarray}
with $q=(x_B p^+,\vec{q}_\perp)$, $x_2=1-x_B-x_1$ and
$\vec{k}_{2\perp}=-\vec{q}_\perp-\vec{k}_{1\perp}$.

\end{document}